\documentclass[
 reprint,
superscriptaddress,
 amsmath,amssymb,
 aps,
superscriptaddress,
 amsmath,amssymb,
 aps,longbibliography
]{revtex4-1}

\usepackage{graphicx}
\usepackage{dcolumn}
\usepackage{bm}
\usepackage{amsmath,amssymb,amstext,amsthm}
\usepackage{braket}
\usepackage{bbold}
\usepackage{bbm}
\usepackage{xcolor}
\usepackage{ gensymb }
\usepackage{soul}
\setstcolor{red}
\usepackage{ mathrsfs }

\begin{document}

\preprint{APS/123-QED}
\setlength{\abovedisplayskip}{1pt}
\title{Experimental Realization of Neutron Helical Waves}

\author{D. Sarenac}
\email{dsarenac@uwaterloo.ca}
\affiliation{Institute for Quantum Computing, University of Waterloo,  Waterloo, ON, Canada, N2L3G1}
\author{M. E. Henderson} 
\affiliation{Institute for Quantum Computing, University of Waterloo,  Waterloo, ON, Canada, N2L3G1}
\affiliation{Department of Physics and Astronomy, University of Waterloo, Waterloo, ON, Canada, N2L3G1}
\author{H. Ekinci} 
\affiliation{Institute for Quantum Computing, University of Waterloo,  Waterloo, ON, Canada, N2L3G1}
\affiliation{Department of Physics and Astronomy, University of Waterloo, Waterloo, ON, Canada, N2L3G1}
\author{Charles W. Clark}
\affiliation{Joint Quantum Institute, National Institute of Standards and Technology and University of Maryland, College Park, Maryland 20742, USA}
\author{D. G. Cory}
\affiliation{Institute for Quantum Computing, University of Waterloo,  Waterloo, ON, Canada, N2L3G1}
\affiliation{Department of Chemistry, University of Waterloo, Waterloo, ON, Canada, N2L3G1}

\author{L. Debeer-Schmitt} 
\affiliation{Neutron Scattering Division, Oak Ridge National Laboratory, Oak Ridge, TN 37831, USA}

\author{M. G. Huber}
\affiliation{National Institute of Standards and Technology, Gaithersburg, Maryland 20899, USA}
\author{C. Kapahi} 
\affiliation{Institute for Quantum Computing, University of Waterloo,  Waterloo, ON, Canada, N2L3G1}
\affiliation{Department of Physics and Astronomy, University of Waterloo, Waterloo, ON, Canada, N2L3G1}

\author{D. A. Pushin}
\email{dmitry.pushin@uwaterloo.ca}
\affiliation{Institute for Quantum Computing, University of Waterloo,  Waterloo, ON, Canada, N2L3G1}
\affiliation{Department of Physics and Astronomy, University of Waterloo, Waterloo, ON, Canada, N2L3G1}

\date{\today}


\pacs{Valid PACS appear here}


\begin{abstract}

Methods of preparation and analysis of structured waves of light, electrons, and atoms have been advancing rapidly. Despite the proven power of neutrons for material characterization and studies of fundamental physics, neutron science has not been able to fully integrate such techniques due to small transverse coherence lengths, the relatively poor resolution of spatial detectors, and low fluence rates. Here, we demonstrate methods that are practical with the existing technologies, and show the experimental achievement of neutron helical wavefronts that carry well-defined orbital angular momentum (OAM) values. We discuss possible applications and extensions to spin-orbit correlations and material characterization techniques.

\end{abstract}
\maketitle

\section{Introduction}

Neutrons play a vital role in characterization of materials and the experimental verification of fundamental physics~\cite{willis2017experimental,Sam}. They are distinguished as a unique probe by their nanometer-sized wavelengths,  magnetic sensitivity, and penetrating abilities stemming from electrical neutrality. Further versatility is enabled by the coupling of neutron's internal and spatial degrees of freedom~\cite{shen2020unveiling}, and the advances in neutron interferometry techniques that exploit the phase degree of freedom of the neutron wavefunction~\cite{pushin2011experimental,shahi2016new}. 


Structured wavefunctions of light, electrons, and atoms have become widely used scientific tools~\cite{rubinsztein2016roadmap,mair2001entanglement,Andersen2006,chen2021engineering,ni2021multidimensional,sarenac2020direct,BarnettBabikerPadgett,ivanov2022promises}. A general class of structured wavefunctions is indexed by orbital angular momentum (OAM)~\cite{Bazhenov1990, LesAllen1992,uchida2010generation,mcmorran2011electron,luski2021vortex}, in which the wavefunction varies as $e^{i\ell\phi}$, where $\ell$ is the OAM value and the angle $\phi$ describes the azimuth around the propagation vector. 
The OAM states possess a ``helical'' or ``twisted'' wavefront, and they have been shown to manifest unique sets of selection rules and scattering cross sections when interacting with matter~\cite{schmiegelow2016transfer,afanasev2018experimental,sherwin2022scattering,afanasev2021elastic,afanasev2019schwinger}. 

It is straightforward to impress OAM upon light with refractive and diffractive optics. However, the full extension of OAM techniques into neutron science has been complicated by several factors.  First, the neutron index of refraction of common materials is on the order of $n\approx1-10^{-5}$. Thus, the basic optimal element, a lens, is not practical in neutron setups. Second, the spatial coherence of neutron beams is at most a few micrometers, as it is typically set by the neutron wavelength, aperture size, and the distance from the aperture to the sample~\cite{sarenac2018three}. The imaging of fine details of diffraction is further hindered by the spatial resolution of neutron cameras that are at best a fraction of a millimeter. Lastly, at powerful research reactors the fluence rate of thermal neutrons at monochromatic [polychromatic] beamlines reaches $10^5$ $[10^7]$ neutrons/(cm$\times$s). For comparison, note that a typical 1 mW red laser emits over $10^{15}$ photon/s.

\begin{figure*}
    \centering\includegraphics[width=\linewidth]{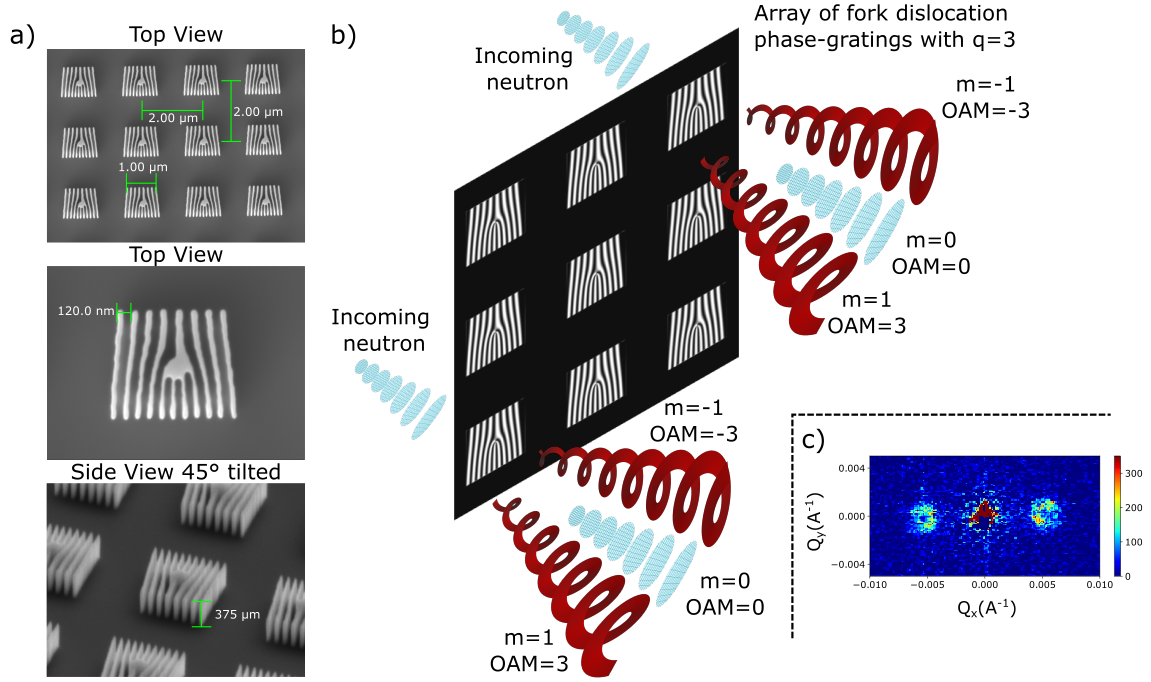}
    \caption{a) SEM images characterizing the array of fork dislocation phase-gratings used to generate the neutron helical wavefronts. The array covered a 0.5 cm by 0.5 cm area and consisted of 6,250,000 individual 1 $\mu m$ by 1 $\mu m$ fork dislocation phase-gratings that possesses a period of $120$ nm, height $500$ nm, and were separated by 1 $\mu m$ on each sides. Two arrays were used in the experiment, differentiated by the fact that one possessed a topology $q=3$ (shown here) and the other $q=7$. 
    b) Each phase-grating with a fork dislocation generates a diffraction spectra consisting of diffraction orders ($m$) that carry a well-defined OAM value of $\ell=m\hbar q$. c) The intensity in the far-field is the sum over the signal from all of the individual fork dislocation phase-gratings. Shown is an example of the collected SANS data.
    }
    \label{fig:concept}
\end{figure*}

Several experimental demonstrations with neutron OAM have been presented~\cite{Dima2015,sarenac2016holography,sarenac2019generation}, reporting the first manipulation of the average neutron OAM value using interferometric measurements of relative phase shifts. However, the creation of a neutron state with helical wavefronts dominated by a single OAM value remained elusive. The prospect of a wide range of significant applications ushered forth diverse proposals and a lively discussion concerning the possibilities of generating and measuring quantized neutron OAM given the technical challenges \cite{cappelletti2018intrinsic,larocque2018twisting,geerits2021twisting,sarenac2018methods,nsofini2016spin,majkrzak2019coherence,jach2021definitive}. 

Here we demonstrate a holographic approach to tuning of neutron OAM. We use microfabricated arrays of millions of diffraction gratings with critical dimensions comparable to neutron coherence lengths and we follow the evolution of the neutron wave packets from micrometer to decimeter length scales. The arrays can be laid out on the cm-square areas typical of usual neutron scattering targets. This suggests possibilities for the direct integration of other structured wave techniques, such as the generation of Airy and Bessel beams~\cite{Harris2015,berry1979nonspreading,durnin1987diffraction}, into neutron sciences. Furthermore, we discuss the applications towards characterization of materials, helical neutron interactions, and spin-obit correlations. 

\section{Preparation and Detection of neutron helical waves}

Phase-gratings with $q$-fold fork dislocations are a standard tool in optics that produce photons with OAM value of $\ell=m\hbar q$ at the m$^{th}$ order of diffraction~\cite{heckenberg1992laser}. This requires that the transverse coherence length of the light beam be at least comparable to the dimensions of the fork dislocation.

Neutron beams have transverse coherence lengths of microns and fluence rates of $10^5-10^7$ neutrons/(cm$^2\times$s). Observing the neutron signal from a single micron sized target is impractical. However, we can multiply the signal by using an $N\times N$ array of micron-sized fork dislocation gratings. When considering an array it is important that the overall array size is much smaller than the diffraction signal of interest, and that the separation distance between the individual gratings is large enough so that it does not induce an observable diffraction order. 

We have fabricated such arrays with $N=2500$ on silicon substrates using electron beam lithography. Fig.~\ref{fig:concept}a shows scanning electron microscope (SEM) images of the fork dislocation phase-gratings with $q=3$. By construction, the spatial dimensions of the individual gratings are comparable to the transverse coherence length of our neutron beam. The use of such an array increases the neutron intensity by $N^2$ in a given $m>0$ diffraction order in the far-field (see Fig.~\ref{fig:concept}b). 
The individual diffraction orders in the presented intensity profiles (see Fig.~\ref{fig:concept}c) span an area of $\approx 10$ cm by $10$ cm, were taken over a period of $\approx40$ min, and consist of the signal from $6,250,000$ individual fork dislocation phase-gratings.

To characterize the generated OAM states we can map out the momentum distribution. Small Angle Neutron Scattering (SANS) beamlines provide several advantages as they map the spatial profiles in the far-field, where the observed intensity distribution is directly determined by the Fourier transform of the outgoing neutron wavefunction. Having access to the far-field enables the use of holographic techniques that have been developed for optical structured waves~\cite{heckenberg1992laser}. Another advantage is the relatively large flux and the accessibility to a wide range of wavelengths. And lastly, it is the typical setup used in material characterization techniques including the contemporary techniques analyzing skyrmion and topological geometries~\cite{henderson2021characterization,henderson2021skyrmion}. Straight-forward extensions follow for incorporating the characterization of materials and performing experiments with helical neutron interactions. 

\begin{figure*}
    \centering\includegraphics[width=\linewidth]{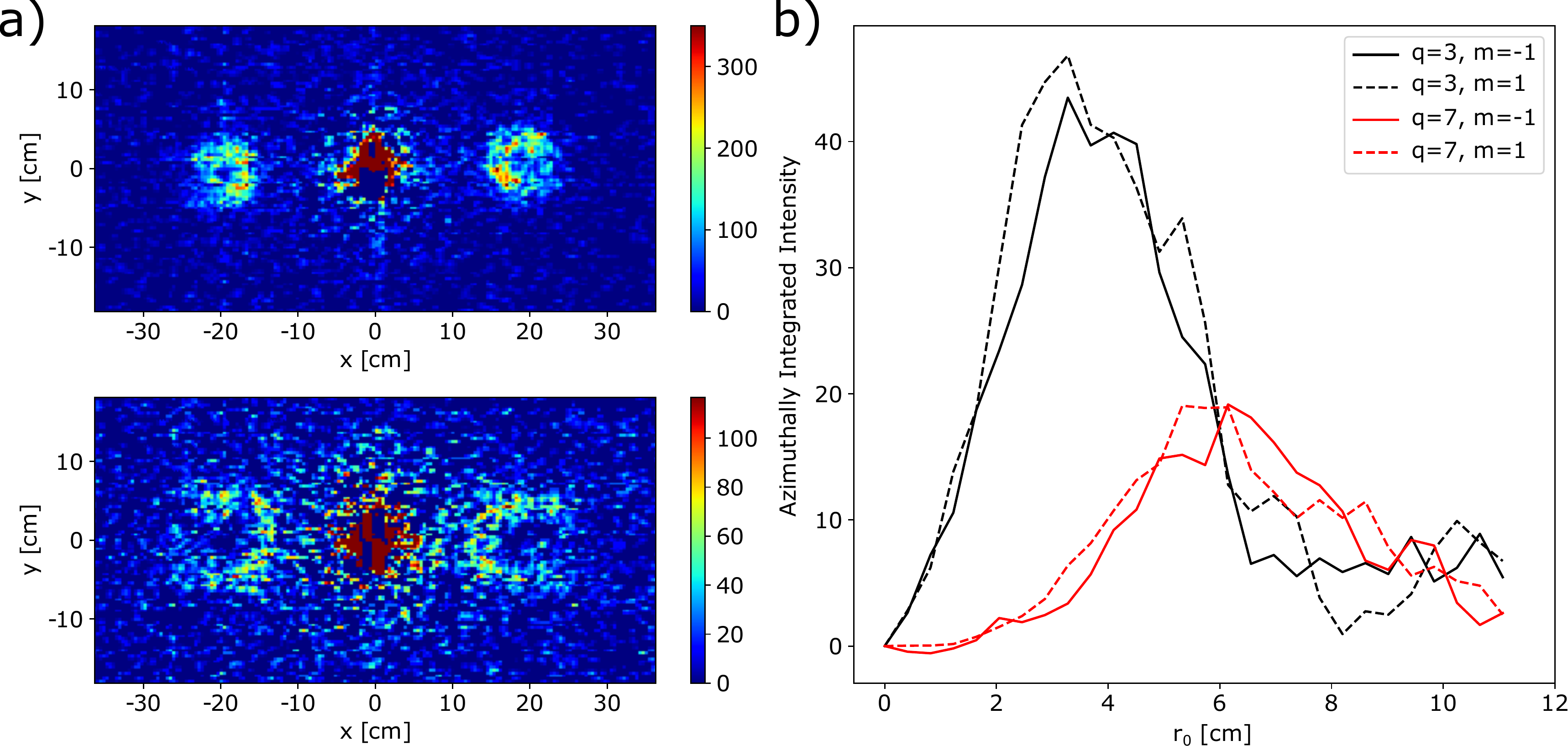}
    \caption{a) SANS data collected for the two arrays of fork dislocation phase-gratings with q=3 (top) and q=7 (bottom). The $(-1,0,1)$ diffraction orders are visible, and the corresponding doughnut profiles induced by the helical wavefronts (see Fig.~\ref{fig:concept}b) can be observed. The measurement time for q=3 was 40 min and for the q=7 was 60 min. b) The azimuthally integrated intensity profiles across the diffraction orders characterizing the size of the doughnut profiles. The point $r_0=0$ corresponds to the center of the diffraction order, and the location of the expected peak of the doughnut profile is given by Eq.~\ref{eqn:rmax}. 
    }
    \label{fig:results}
\end{figure*}

A simple map for modelling the action of a binary phase-grating with a fork dislocation can be expressed as:

\begin{align}
\psi_\text{in}\rightarrow\psi_\text{in}\left[\cos\left(\frac{\alpha}{2}\right)+\sin\left(\frac{\alpha}{2}\right)\sum_m \frac{2}{m\pi}e^{i\frac{2\pi mx}{p}}e^{imq\phi}\right]
\label{Eqn:gratingMap}
\end{align}

\noindent where where $x$ ($\phi$) is the Cartesian (azimuthal) coordinate, $p$ is the grating period, $m=...-3,-1,1,3...$ are the non-zero diffraction orders, $\alpha$ is the induced phase by the height of the grating grooves (see Appendix A), and the incoming wavefunction $\psi_\text{in}$ is typically taken to be a Gaussian profile for convenience.  The far-field is typically defined to be the distance at which the diffraction orders are spatially separated, so here we can consider the $m$ terms independently along their respective propagation directions. We thus obtain well-defined OAM states in the form of $\psi_\text{in}e^{imq\phi}$. The full analysis of the evolution of such states is presented in Ref.~\cite{berry2004optical}. Consider one diffraction order of a single neutron wavepacket that transverses on-axis with the fork dislocation of a single phase-grating. Each point ($r_\mathrm{0},\phi_\mathrm{0}$) in the neutron wavepacket is diffracted along the transverse direction, such that the induced OAM relative to the axis of propagation is independent of the point's location~\cite{berry1998paraxial}: 

\begin{align}
\ell=|\vec{r}\times\vec{p}|=r_\mathrm{0}\hbar k_\perp=\hbar q,
\end{align}

\noindent where $\vec{p}$ is the neutron momentum, $k_\perp$ is the magnitude of the wavevector along the $\phi$ direction, and $\hbar$ is the reduced Planck's constant. Therefore, every part of the neutron wave packet obtains a well-defined OAM of $\ell=\hbar q$. 

With equal transverse coherence lengths $\sigma_x=\sigma_y=\sigma_\perp$ we can make use of the cylindrical symmetry to describe the transverse wave function in terms of solutions to the 2-D harmonic oscillator~\cite{nsofini2016spin}:

\begin{align}
\psi_{\ell,n_r}(r,\phi)=\mathcal{N}\left(\frac{r}{\sigma_\perp}\right)^{|\ell|}e^{-\frac{r^2}{2\sigma^2_\perp}}\mathcal{L}_{n_r}^{|\ell|}\left(\frac{r^2}{\sigma^2_\perp}\right)e^{i\ell\phi},
\end{align}

\noindent where  $\mathcal{N}=\frac{1}{\sigma_\perp}\sqrt{\frac{n_r !}{\pi(n_r+|\ell|)!}}$ is the normalization constant, $n_r\in(0,1,2...)$, $\ell\in (0, \pm 1, \pm2...) $, and $\mathcal{L}_{n_r}^{|\ell|}\left(r^2/\sigma^2_\perp\right)$ are the associated Laguerre polynomials. The corresponding neutron energy is

\begin{align}
E =\hbar\omega_\perp(2n_r+|\ell|+1),
\end{align}  

\noindent where $\omega_{\perp}^2=\hbar/(2M\sigma_{\perp}^2)$, and $M$ is the mass of the neutron. Each diffraction order  $m$ of the fork dislocation phase-grating is in a definite state of OAM:

\begin{align}
\psi=\sum_{n_r}\psi_{\ell=mq,n_r}.
\end{align}  

Considering $n_r=0$ dominant term~\cite{nsofini2016spin} of the first diffraction order, we can determine that the azimuthally integrated intensity:

\begin{align}
\int_0^{2\pi}|\psi_{q,0}(r_0,\phi_0)|^2d\phi
\label{eqn:azimuthalIntensity}
\end{align}  

\noindent peaks at:

\begin{align}
r_0 =\sigma_\perp\sqrt{q}.
\label{eqn:rmax}
\end{align}

\section{Results and Analysis}

The experiments were performed on the GP-SANS beamline at the High Flux Isotope Reactor at Oak Ridge National Laboratory~\cite{wignall201240}. The experimental setup parameters are described in detail in Appendix B. A circular aperture of $s=20$ mm diameter is used to define symmetric transverse coherence lengths $\sigma_x=\sigma_y=\sigma_\perp$. With the aperture to sample distance of $L_1=17.8$ m and neutron wavelength of $\lambda=12$~\AA~we set the transverse coherence of the neutron wavepacket to the size of one fork dislocation phase-grating $\sigma_\perp\approx\lambda L_1/s\approx1$ $\mu m$. The distance between the sample and the camera was 19 m, and the camera size span an area of $\approx 1$ $m^2$ with each pixel being $\approx 5.5$ mm by $4.1$ mm in size.

We fabricated two arrays of fork dislocation phase-gratings on Si wafers, one with a $q=7$ topology and the other with $q=3$. Each array covered a 0.5 cm by 0.5 cm area and consisted of $6,250,000$ individual 1 $\mu m$ by 1 $\mu m$ fork dislocation phase-gratings, where each one possessed a period of $120$ nm, height $500$ nm, and was separated by 1 $\mu m$ on each side from the other fork dislocation phase-gratings. The full consideration for the parameters is given in the Appendix A as well as the fab procedure in Appendix C.

The observed SANS data for the two arrays of fork dislocation phase-gratings is shown in Fig.~\ref{fig:results}a. With the phase-grating period of $p=120$ nm, the angle of divergence of the first diffraction order is $\theta\approx\lambda/p\approx0.01$ rad which corresponds to $Q_x=2\pi/p=0.00525$ on the SANS images (see Fig~\ref{fig:concept}c). In our particular setup this corresponds to a spatial distance of $x=19$ cm on the camera as shown in Fig.~\ref{fig:results}a. Good agreement is found with the observed location of the peaks.

To quantify the doughnut profiles we can analyze the azimuthally integrated intensity (Eq.~\ref{eqn:azimuthalIntensity}) centered on the first diffraction orders. As per Eq.~\ref{eqn:rmax} the expected doughnut size varies with $\sqrt{q}$, which is in good agreement with the observed data.

\section{Discussion and Conclusion}

We have described and experimentally demonstrated a method to generate and characterize neutron helical waves that are dominated by a well-defined OAM value. The experimental demonstration consisted of fabricating a 2D array of fork dislocation phase-gratings. With a pattern area of 0.5 cm by 0.5 cm that amounts to 6,250,000 individual fork dislocation phase-gratings contributing to the observed signal. $q=3$ and $q=7$ topologies were tested and the obtained doughnut shaped intensity distributions are in good agreement with theory.

The presented method opens the door for the implementation of other structured wave techniques with neutrons, such as the generation of ``self-accelerating'' Airy beams~\cite{berry1979nonspreading,siviloglou2007observation}, as well as the ``non-diffractive'' Bessel beams~\cite{durnin1987diffraction,garces2002simultaneous}. These beams possess a ``self-healing'' property as they can reform after being partially obstructed. Considering an array of phase-gratings, the Airy beams would be generated through the addition of a cubic phase gradient while the Bessel beams through the addition of a radial phase gradient.

The convenient integration with a SANS beamline provides access to the far-field where the helical beams with specific OAM values are separated in the form of diffraction orders. This enables studies of  interactions between neutron's helical degree of freedom and scattering from materials. For example, placing the fork array phase-grating before a topological sample ~\cite{henderson2021characterization,henderson2021skyrmion} and post-selecting the analysis on individual diffraction orders allows for direct study of scattering properties from neutron helical waves.

Another avenue of exploration that is made possible is 
the experimental investigation of neutron selection rules. For example, this may be achieved through the addition of $^3$He spin filters. The absorption of neutrons by nuclear spin-polarized $^3$He is strongly dependant on the spin orientation of the neutron due to the conservation of spin angular momentum. Our method allows for the direct tests at a SANS beamline through the characterization of the diffraction peaks after the post-selection via $^3$He cell polarizers.

\section*{Acknowledgements}

This work was supported by the Canadian Excellence Research Chairs (CERC) program, the Natural Sciences and Engineering Research Council of Canada (NSERC), the Canada  First  Research  Excellence  Fund  (CFREF), and the US Department of Energy, Office of Nuclear Physics, under Interagency Agreement 89243019SSC000025.

\bibliography{OAM}

\clearpage

\thispagestyle{empty}
\section*{APPENDIX}

\subsection{Fork Array Parameters}

A $50\%$ duty cycle phase-grating with a fork dislocation has the following profile:

\begin{align}
	F(x)=\frac{\alpha}{2} \text{sign}\left(\text{sin}\left[\frac{2\pi  }{p}x+q\phi\right]\right),
	\label{eqn:gratingProfile}
\end{align}

\noindent where $p$ is grating period, $q$ is the topological charge, $x$ is the Cartesian coordinate, and $\phi$ is the azimuthal coordinate. The groove height $h$ sets the phase shift that is induced by each grating groove: $\alpha=-Nb_c\lambda h$ where $Nb_c$ is the coherent scattering length density of the grating material and $\lambda$ is the neutron wavelength. As per Eq.~\ref{Eqn:gratingMap} we can see that $\alpha$ determines the relative amplitudes of the diffraction orders. For example, for $\alpha=\pi$ the zeroth order is suppressed while for $\alpha=\pi/2$ there is an equal amount of zeroth order and the higher orders. Note that the fabrication of the small periods currently limits us to small $\alpha$, and hence we will not be considering these effects. The only consideration is given to minimizing the acquisition time by maximizing the height $h$ to increase the number of neutrons in the first diffraction order. 

The resulting profile for $q=3$ is shown in Fig.~\ref{fig:concept}a. The period $p$ determines the angle of propagation of the diffraction orders, and hence it needs to be set by the requirements of the given beamline. The angle of propagation of the first diffraction order is: 

\begin{align}
	\theta=\text{sin}^{-1}\left(\frac{\lambda}{p}\right)\approx\frac{\lambda}{p}\approx\frac{\lambda Q_G}{2\pi}
	\label{eqn:diffAngle}
\end{align}
\noindent where $Q_G=2\pi/p$ is the scattering vector of the gratings.

The topological charge $q$ sets the OAM values carried by the diffraction orders. As depicted on Fig.~\ref{fig:concept}a, $q$ is the difference between the number of periods along the grating direction above the origin when compared to the number below the origin. 

It is important to note that the fabrication challenges of the central region of the ideal fork dislocation phase-grating are typically overcome through the omission of the central region. Accordingly, a flat circular profile with diameter $\approx200$ nm was intentionally imposed in our designs, which can be observed in the SEM figures. This is a common practice in optical OAM techniques given that the  effects of such a feature are negligible in diffraction as the beam diverges away from the center.  

Lastly the purpose of creating an array of these structures is two-fold. The first is that by creating the array with identical copies we are able to increase the signal to measurable amounts.  Note that the spacing between the individual fork dislocation phase-gratings is $1$ $\mu m$ over the $0.5$ cm by $0.5$ cm area, whereas the diameter of the ring in the first diffraction order is several centimeters. Hence the measured signal is the integral over the positions of the individual gratings whose separation distance and size is much smaller than the signal of interest. The $q=3$ images shown in Fig 2 were taken over 40 min and consist of the signal from millions of fork dislocation phase-gratings. Therefore, it is not practical to measure the signal of a single such grating. The second advantage is that future studies with materials are conveniently integratable with this design. A topological material typically possesses an array of topologies and hence it is desirable to create a tool with similar properties for more specific characterization. 

\subsection{Experimental Setup}

Small-angle neutron scattering measurements were performed on the GP-SANS beamline at the High Flux Isotope Reactor at Oak Ridge National Laboratory~\cite{wignall201240}. The fork dislocation phase-gratings were placed inside a ThorLabs rotation mount (pat number RSP2D) which was then affixed to the end of the sample aperture holder. The gratings were placed 17.8 m away from the 20 mm diameter source aperture. A 4 mm diameter sample aperture was placed right in front of the sample to reduce possible background. The distance from the grating to the camera was 19 m, and a wavelength distribution of $\delta\lambda/\lambda=10\%$ and a central wavelength of $12$~\AA . The resulting standard deviation of the resolution distribution was estimated as $\sigma_{Qx}$=0.0001645~\AA. 

\subsection{Fork Array Microfabrication and Characterization}

The target parameters for the fork dislocation phase-gratings were: period: $p=120$ nm for both $q=3$ and $q=7$, groove height of $h=500$ ($400)$ nm for $q=3$ ($q=7$), and inner region diameter $200$ nm for both $q=3$ and $q=7$. These parameters were experimentally optimized to ensure that we obtain high quality and robust structures. 

Double-side polished, intrinsic, 2 inch diameter (100) silicon wafers were used to fabricate the arrays of fork dislocation phase-gratings. Electron beam lithography (EBL) was employed to pattern the high performance positive EB resists (ZEP520A, $\approx$ 80 nm). The e-beam exposure was carried out with a JEOL JBX-6300FS EBL system operating at 100 kV and 2 nA beam current. The e-beam dosage was 250 $\mu$C/cm$^2$. After the e-beam exposure, the sample was processed in the developer ZED-N50 for 90 s, and then immersed in IPA for 60 s followed by a pure nitrogen dry. As hard mask during plasma etching of Si, Cr metal (20 nm) was e-beam evaporated and lifted-off in heated PG Remover. A pseudo-Bosch recipe was adopted to achieve vertical sidewall etch profile. The samples were etched in an Oxford PlasmaLab ICP-380 inductively coupled plasma reactive ion etching (ICP-RIE) system, which provides high-density plasma with independently controlled system parameters. The recipe includes rf: 100 W, ICP:1200 W, C$_4$F$_8$: 25 sccm, SF$_6$: 15 sccm, pressure: 10 mTorr, temperature: 20$\degree$C. After fabricating the array of fork dislocation phase-gratings, the remaining Cr etch mask was removed via plasma etching.

\begin{figure*}
    \centering\includegraphics[width=0.75\linewidth]{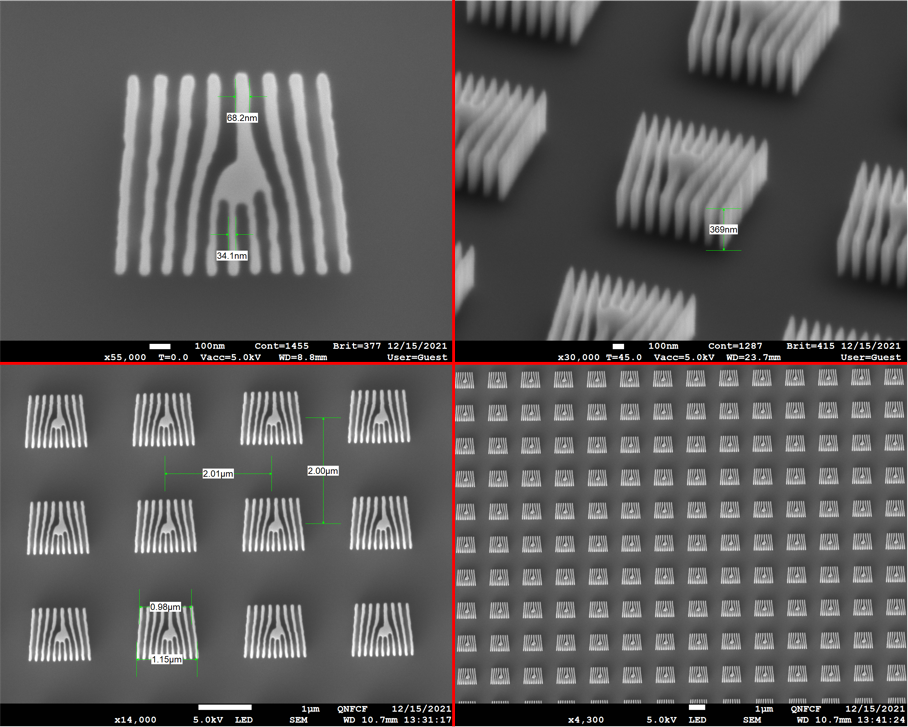}
    \caption{ SEM characterization of the 2D fork dislocation phase-grating array with topology q =3. Shown images are top view except the top right image that is at the $45\degree$ tilt. The array covered a 0.5 cm by 0.5 cm area and consisted of 6,250,000 individual 1 $\mu m$ by 1 $\mu m$ fork dislocation phase-gratings, where each grating possessed a period of $120$ nm, height $500$ nm, and was separated by 1 $\mu m$ on each side from the other gratings.
    }
    \label{fig:fab_fig_oam3}
\end{figure*}

\begin{figure*}
    \centering\includegraphics[width=0.75\linewidth]{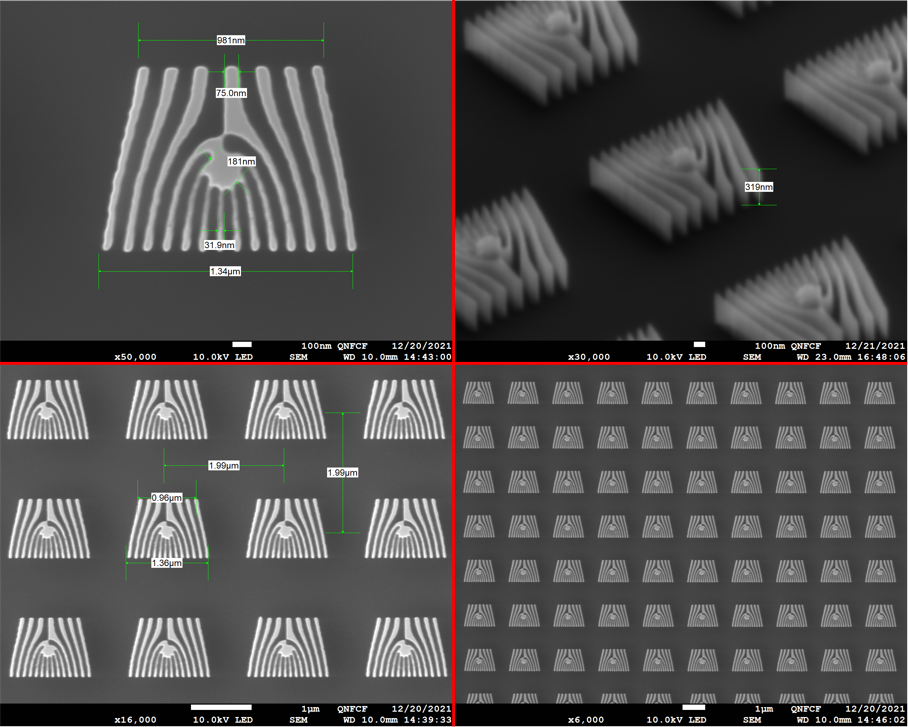}
    \caption{ SEM characterization of the 2D fork dislocation phase-grating array with topology q =7. Shown images are top view except the top right image that is at the $45\degree$ tilt. The array covered a 0.5 cm by 0.5 cm area and consisted of 6,250,000 individual 1 $\mu m$ by 1 $\mu m$ fork dislocation phase-gratings, where each grating possessed a period of $120$ nm, height $400$ nm, and was separated by 1 $\mu m$ on each side from the other gratings.
    }
    \label{fig:fab_fig_oam3}
\end{figure*}

\end{document}